\documentclass[aps,prd,onecolumn,amsmath,11pt,superscriptaddress,floatfix,nofootinbib,preprintnumbers]{revtex4-2}

\usepackage{mathrsfs}
\usepackage{amsfonts}
\usepackage{amsmath,amssymb,bm}
\usepackage{array}
\usepackage{verbatim}
\usepackage{epsfig}
\usepackage{graphicx}
\usepackage{hyperref}
\hypersetup{colorlinks, linkcolor = [rgb]{0, 0, 0.5}, citecolor = [rgb]{0,0.0,0.5}, urlcolor = [rgb]{0,0.0,0.5}}
\usepackage[normalem]{ulem}
\usepackage{xcolor}
\usepackage{ulem}
\usepackage{multirow}

\usepackage{subcaption}
\usepackage{graphicx}
\graphicspath{{./figs/}}
\usepackage{dcolumn}
\usepackage{bm}
\usepackage{slashed}
\usepackage{siunitx}
\usepackage{ulem,xpatch}
\usepackage{hyperref}
\usepackage[mathlines]{lineno}
\usepackage{comment}
\usepackage{soul}
\usepackage{xcolor}
\usepackage{xfrac}

\allowdisplaybreaks[4]

\begin{document}

\title{Study the impact of polarized background fields on coupling constants in EIC and EicC}

\author{Cong Li}
\affiliation{School of Information Engineering, Zhejiang Ocean University, Zhoushan, Zhejiang, China}
\author{Jing Zhao}
\affiliation{Key Laboratory of Particle Physics and Particle Irradiation (MOE), Institute of Frontier and Interdisciplinary Science, Shandong University, Qingdao, Shandong 266237, China}
%


\begin{abstract}
In the polarized background field, the coupling constant will be influenced. We quantify this effect and propose that it can be measured at the EIC and EicC through the Bethe-Heitler process.
\end{abstract}

\maketitle
\section{Introduction}
\label{s.intro}
Recent experimental observations reveal that background fields significantly influence the coupling constant. 
As is well-known, the Color Glass Condensate (CGC) \cite{CGC1,CGC2,CGC3} and Quark-Gluon Plasma (QGP) \cite{QGP1,QGP2,QGP3,QGP4,QGP5} cannot be adequately described by perturbative quantum chromodynamics (pQCD). 
The reason lies in the unique features of these systems: both are characterized by high-density gluon fields, which substantially modify the nature of interactions. Notably, there exists a positive correlation between high gluon density and an anomalous enhancement of the coupling constant. Studying the influence of background fields on the coupling constant is crucial for deepening our understanding of QCD theory and holds significant value for the precise experimental measurement of QCD. In experimental studies of QCD, due to the color confinement, hadrons or nuclei often act not only as the interacting entities but also as effective background fields, which in turn modify the coupling constants in these processes. 

When hadrons or nuclei serve as background fields, the gluons within them may exhibit pronounced linear polarization \cite{crosssection2,lg,lg2}. To explore this phenomenon, we draw an analogy to quantum electrodynamics (QED), where photons play a role analogous to gluons in QCD. In our earlier research \cite{Li1,Li2}, we analyzed the photon distribution in ultra-peripheral collisions (UPCs), parameterizing photon correlation functions to demonstrate, in theory, that photons can exhibit significant linear polarization. We proposed that this polarization could be experimentally verified by measuring the azimuthal asymmetries of final-state particles. Subsequent STAR collaboration \cite{star} affirmed our theoretical prediction, confirming the polarized nature of UPC photons. Since QED and QCD share structural similarities, this result motivates further investigation into polarized gluon fields in QCD. Then, it is feasible to study the effects of polarized photon background fields on the electromagnetic coupling constant by considering QED processes in the photon fields surrounding atomic nuclei.

Additionally, Ref.~\cite{cc3} proposed exploring Coulomb corrections in the Bethe-Heitler (BH) process during electron-nucleus (e-A) collisions at future facilities such as the Electron-Ion Collider (EIC) in the United States and the Electron-Ion Collider in China (EicC). Building upon this work, our study incorporates three key aspects: (1) the selection of kinematic regions relevant to the problem, (2) the inclusion of Coulomb corrections, and (3) the role of Sudakov factors arising from soft photon radiation in the final state.
Specifically, we focus on the BH process, in which photons and electrons interact to produce photons and electrons. In this process, final-state photon radiation occurs within the external electric field of the nucleus, which, as mentioned above, is polarized. By measuring the azimuthal asymmetries of the final-state particles in this scenario, we aim to quantify how the polarized photon field affects the electromagnetic coupling constant. These studies are expected to offer insights into analogous QCD dynamics, particularly in relation to polarized gluon fields.

The structure of this paper is organized as follows. In Sec.\ref{2}, we give the correction of the coupling constants arising from the polarized background fields. In Sec.\ref{s.observables}, we derive the cross-section for the BH process, incorporating the influences of polarized background fields, Coulomb corrections, and Sudakov factors. Furthermore, we provide numerical evaluations that correspond to our theoretical predictions. In Sec.~\ref{4}, we summarize our findings and their implications for future studies at the EIC and EicC.

\section{Correction of coupling constant by background field}
\label{2}

Considering the BH process in electron-nucleus (eA) collisions at the EIC and EicC, one can apply the transverse momentum dependent (TMD) factorization~\cite{Collins:1981uw,Collins:1981uk,Aybat:2011zv} to express the cross section in terms of the convolution of the hard scattering part and the photon TMD distributions.
When an electron traverses a photon field, it undergoes multiple scatterings with the background photon field, and the cumulative effect of these scatterings is referred to as the Coulomb correction.
This process can occur either before the emission of a photon or after a photon has been radiated.
The Coulomb correction can be incorporated into the photon correlation functions in the form of a closed-loop gauge link \cite{cc1,cc2,cc3}, expressed as
\begin{align}
&\int\frac{dy^-d^2y_\bot}{P^+(2\pi)^3}e^{ik\cdot y} \langle A|F_{+\bot}^\mu(0)U^\dag(0_\bot)U(y_\bot)F_{+\bot}^\nu(y)|A\rangle \bigg|_{y^+=0}\nonumber\\
=& \frac{\delta_\bot^{\mu\nu}}{2} x f_1^\gamma(x, k_\bot^2) + \left( \frac{k_\bot^\mu k_\bot^\nu}{k_\bot^2} - \frac{\delta_\bot^{\mu\nu}}{2} \right) x h_1^{\bot\gamma}(x, k_\bot^2), 
\end{align}
where we have the transverse tensor $\delta_\perp^{\mu\nu} = -g^{\mu\nu} + p^{\mu} n^{\nu} + p^{\nu} n^{\mu}$ and $k_\perp^2 = \delta_\perp^{\mu\nu} k_{\perp \mu} k_{\perp \nu}$. The TMD functions $f_1^\gamma(x, k_\perp^2)$ and $h_1^{\perp \gamma}(x, k_\perp^2)$ correspond to the unpolarized and linearly polarized photon distributions, respectively.
The transverse gauge link $U^\dagger(0_\perp)U(y_\perp)$ effectively encapsulates the Coulomb correction of the hard part into the photon TMD distributions. It can be expressed as
\begin{equation}
U\left(y_\bot\right)=\mathcal{P}e^{ie\int_{-\infty}^{+\infty} d\ z^-A^+\left(z^-,y_\bot\right)}.
\end{equation}
We should note that, since the photon is electrically neutral, the correlation function does not require a gauge link to ensure gauge invariance. The gauge potential here can be further represented as
\begin{equation}
\mathcal{V}\left(y_\bot\right)\equiv e\int_{-\infty}^{+\infty} dz^-A^+\left(z^-,y_\bot\right)=\frac{\alpha Z}{\pi}\int d^2q_\bot e^{-iy_\bot\cdot q_\bot}\frac{F\left(q_\bot^2\right)}{q_\bot^2+\delta^2},
\end{equation}
where $\delta$ is the mass of the photon, used to regulate the infrared divergence, and $F$ is the nuclear charge form factor. In our numerical evaluations, the nuclear charge form factor can be obtained from the STARlight MC generator~\cite{starlight},
\begin{equation}
F(|\vec{k}|)=\frac{4\pi\rho^0}{|\vec{k}|^3A}\left[\sin\left(|\vec{k}|R_A\right)-|\vec{k}|R_A\cos\left(|\vec{k}|R_A\right)\right]\frac{1}{a^2{\vec{k}}^2+1},\label{e.mc}
\end{equation}
where $R_A = 1.1 A^{1/3} \, \text{fm}$ and $a = 0.7 \, \text{fm}$. This parametrization closely resembles the Woods-Saxon distribution in phase-space. The form of the field strength in the correlation function is
\begin{equation}
\mathcal{F}^\mu\left(x,y_\bot\right)\equiv\int_{-\infty}^{+\infty} dy^-e^{ixP^+y^-}F_{+\bot}^\mu\left(y^-,y_\bot\right)=\frac{Ze}{4\pi^2}\int\ d^2q_\bot e^{-iy_\bot\cdot q_\bot}\left(iq_\bot^\mu\right)\frac{F\left(q_\bot^2+x^2M_p^2\right)}{q_\bot^2+x^2M_p^2},
\end{equation}
where $x$ is the longitudinal momentum fraction carried by the photon and $M_p$ is the proton mass.

With all the aforementioned components, the correlation function of the photon can be rewritten as
\begin{align}
&\int\ \frac{d^2y_\bot d^2y_\bot^\prime}{4\pi^3}e^{ik_\bot\cdot\left(y_\bot-y_\bot^\prime\right)}\mathcal{F}^\mu\left(x,y_\bot\right)\mathcal{F}^{\ast\nu}\left(x,y_\bot^\prime\right)e^{i\left[\mathcal{V}\left(y_\bot\right)-\mathcal{V}\left(y_\bot^\prime\right)\right]}\nonumber\\
=&\frac{\delta_\bot^{\mu\nu}}{2}xf_1^\gamma\left(x,k_\bot^2\right)+\left(\frac{k_\bot^\mu k_\bot^\nu}{k_\bot^2}-\frac{\delta_\bot^{\mu\nu}}{2}\right)xh_1^{\bot\gamma}\left(x,k_\bot^2\right).
\end{align}
The desired photon TMDs can be obtained by the property of orthogonality and they can be expressed as
\begin{align}
xf_1^\gamma\left(x,k_\bot^2\right)&=\int\ \frac{d^2y_\bot d^2y_\bot^\prime}{4\pi^3}e^{ik_\bot\cdot\left(y_\bot-y_\bot^\prime\right)}\mathcal{F}^\mu\left(x,y_\bot\right)\mathcal{F}^{\ast\nu}\left(x,y_\bot^\prime\right)\delta_\bot^{\mu\nu}e^{i\left[\mathcal{V}\left(y_\bot\right)-\mathcal{V}\left(y_\bot^\prime\right)\right]},\\
xh_1^{\bot\gamma}\left(x,k_\bot^2\right)&=\int\ \frac{d^2y_\bot d^2y_\bot^\prime}{4\pi^3}e^{ik_\bot\cdot\left(y_\bot-y_\bot^\prime\right)}\mathcal{F}^\mu\left(x,y_\bot\right)\mathcal{F}^{\ast\nu}\left(x,y_\bot^\prime\right)\left(\frac{k_\bot^\mu k_\bot^\nu}{k_\bot^2}-\frac{\delta_\bot^{\mu\nu}}{2}\right)e^{i\left[\mathcal{V}\left(y_\bot\right)-\mathcal{V}\left(y_\bot^\prime\right)\right]}.
\end{align}

As mentioned in the Introduction, previous studies~\cite{Li1,Li2} have shown that the peripheral photons of the atomic nucleus are linearly polarized. When regarded as a background field, the propagator of the photon becomes more complex and takes the following form
\begin{equation}
\int \frac{2d^4 k}{(2\pi)} e^{-ik \cdot (x - y)} \frac{-i}{k^2 + i\varepsilon} \left[ g_{\bot \mu \nu} x f_1^\gamma(k_z, k_\bot) - \left( g_{\bot \mu \nu} + 2 \frac{k_{\bot \mu} k_{\bot \nu}}{k_\bot^2} \right) x h_1^{\perp \gamma}(k_z, k_\bot) \right],\label{e.propagator2}
\end{equation}
where $-g_\perp^{\mu\nu} = \delta_\perp^{\mu\nu} = -g^{\mu\nu} + p^{\mu} n^{\nu} + p^{\nu} n^{\mu}$. Note that $\frac{1}{2(2\pi)^3}$ is hidden within the TMDs $f_1^\gamma(k_z, k_\perp^2)$ and $h_1^{\perp \gamma}(k_z, k_\perp^2)$. These TMDs provide both the incident photon and the background field for subsequent reaction processes. In prior studies~\cite{Li3}, we considered a non-polarized background field. The modified propagator in this case is
\begin{equation}
\int{\frac{d^4k}{(2\pi)^4}e^{-ik\cdot\left(x-y\right)}\frac{-i}{k^2+i\varepsilon}}g_{\bot\mu\nu}[{xf}_1^\gamma\left(k_z,k_\bot\right)2(2\pi)^3].\label{e.propagator1}
\end{equation}
It can be decomposed into two components: an uncorrected propagator $\int \frac{d^4 k}{(2\pi)^4} e^{-ik \cdot (x-y)} \frac{-i}{k^2 + i\epsilon} g_{\perp\mu\nu}$ and the function ${xf}_1^\gamma\left(k_z,k_\bot\right)2(2\pi)^3$, where ${xf}_1^\gamma\left(k_z,k_\bot\right)2(2\pi)^3$ can be combined with the coupling constant to form an effective coupling constant
\begin{equation}
\alpha_e^{eff}=\alpha_e{xf}_1^\gamma\left(k_z,k_\bot\right)2(2\pi)^3.
\end{equation}
This can be understood as a modification of the coupling constant due to the influence of a background field. It is evident that a dense background field affects the coupling constant, leading to deviations from perturbative theory.
However, when the background field becomes polarized, the photon propagator becomes complex and cannot be separated into the two independent components as defined in Eq.~\eqref{e.propagator1}. Despite this complexity, the magnitudes of $ f_1^\gamma(k_z, k_\perp^2) $ and $ h_1^{\perp \gamma}(k_z, k_\perp^2) $ will ultimately alter the coupling constant.
In summary, the presence of a polarized background field modifies the photon propagator, impacting the effective coupling constant in a way that diverges from the standard perturbative approach.
Their specific forms can be derived from the equivalent photon approximation.
Moreover, the studies \cite{Li1,Li2,eq,eq2} revealed that the photon distributions $f_1^\gamma(k_z, k_\perp^2)$ and $h_1^{\perp \gamma}(k_z, k_\perp^2)$ are equal outside the nucleon, which gives us
\begin{equation}
xf_1^\gamma\left(k_z,k_\bot^2\right)=xh_1^{\bot\gamma}\left(k_z,k_\bot^2\right)=\frac{Z^2\alpha_e}{\pi^2}k_\bot^2\left[\frac{F\left(k_\bot^2+k_z^2\right)}{\left(k_\bot^2+k_z^2\right)}\right]^2,
\end{equation}
where $Z$ is the nuclear charge number and $F$ is the nuclear charge form factor. The form factor $F$ has been given in Eq.~\eqref{e.mc}.

\section{Observables}
\label{s.observables}
We can study the effect of polarized background fields on coupling constants through the BH process at the EIC and EicC, i.e.,
\begin{equation}
e(k_1)+\gamma(k_2)\longrightarrow e(p_1)+\gamma(p_2),
\end{equation}
where the variables in parentheses indicate the four momenta of the corresponding particles.
In this process, the electron scatters off the atomic nucleus, absorbing a photon from the electric field surrounding the nucleus and subsequently emitting another photon.
Since the emission of the photon occurs in the outer electric field of the nucleus, the corresponding final-state photon propagator can be replaced by Eq.~\eqref{e.propagator2}.
The influence of the polarized background field on the coupling constant can be studied by measuring the azimuthal distribution of the final-state particles.
A \(\cos 2\phi\) azimuthal modulation in the BH cross section can arise from two sources. Firstly, it may result from the initial-state polarization of the photon beam. Secondly, it can also originate from the effect of the background field on the coupling constant when a virtual photon is emitted in the final state, rather than a real one \cite{crosssection1,crosssection2}.
The resummation of all large logarithmic terms leads to the Sudakov factor $\exp{[-\frac{\alpha_e}{2\pi}\ln^2\left(\frac{P_\bot^2}{\mu_r^2}\right)]}$ where $\mu_r = \frac{2e^{-\gamma_E}}{\left|r_\bot\right|}$. Upon convoluting the Sudakov factor in the collision parameter space, we have
\begin{align}
        \frac{d\sigma}{dy_\gamma d^2P_\bot d^2q_\bot} = & 2(2\pi)^3\int \frac{d^2r_\perp}{(2\pi)^2} e^{ir_\bot \cdot q_\bot} e^{-\frac{\alpha_e}{2\pi} \ln^2 \frac{Q^2}{\mu_r^2}}\nonumber \\
    & \times \int d^2k_\bot \, e^{ir_\bot \cdot k_\bot} \bigg[ H_1 x f_1^\gamma(x, k_{2\bot}^2) xf_1^\gamma(p_{2z}, p_{2\bot}^2)+ H_2 x h_1^{\bot\gamma}(x, k_{2\bot}^2) xf_1^\gamma(p_{2z}, p_{2\bot}^2) \nonumber\\
    &+ H_3 x f_1^\gamma(x, k_{2\bot}^2) xh_1^{\bot\gamma}(p_{2z}, p_{2\bot}^2)+ H_4 x h_1^{\bot\gamma}(x, k_{2\bot}^2) xh_1^{\bot\gamma}(p_{2z}, p_{2\bot}^2) \bigg],
\end{align}
where
\begin{align}
    &H_1=\frac{2z^2(P_\bot^4(3z^2-4z+2)-2P_\bot^2Q^2(z-2){(z-1)}^2+Q^4{(z-1)}^2(z^2-2z+2))}{{(P_\bot^2-Q^2z+Q^2)}^4},\\
&H_2=\frac{4P_\bot^2(z-1)z^3(P_\bot^2+Q^2)}{{(P_\bot^2-Q^2z+Q^2)}^4}\cos{2\phi},\\
&H_3=\frac{4P_\bot^2\ \left(z-1\right)z^3}{\left(P_\bot^2-Q^2z+Q^2\right)^3},\\
&H_4=\frac{4{(z-1)}^2z^2(P_\bot^2+Q^2)}{{(P_\bot^2-Q^2z+Q^2)}^3}\cos{2\phi},
\end{align}
and $z$ represents the longitudinal momentum fraction of the incoming electron carried by the final-state photon. The total transverse momentum $q_{\perp} = k_{2\bot} = p_{1\bot} + p_{2\bot} \approx 0$ is quite small, thus $p_{1\bot} \approx -p_{2\bot} \approx P_\bot$. Here, $Q^2$ denotes the virtuality of the final-state virtual photon, and $\phi$ is the angle between $P_\bot$ and the total transverse momentum $k_{2\bot}$. Interestingly, the $H_3$ term does not contain $\cos{2\phi}$ which is canceled during the derivation. Similarly, the $\cos{4\phi}$ term originally present in $H_4$ also disappears, leaving only the $\cos{2\phi}$ term. Therefore, the azimuthal asymmetry of $\cos{4\phi}$ cannot observed in experiments. However, this does not imply that the effect of the background field on the coupling constant is absent. 
For comparison, one can obtain the cross section of the same process without considering the background field influence directly from the literature \cite{cc3}
\begin{align}
    \frac{d\sigma}{dy_\gamma d^2P_\bot d^2q_\bot} = &\int \frac{d^2r_\bot}{(2\pi)^2} e^{ir_\bot \cdot q_\bot} e^{-\frac{\alpha_e}{2\pi} \ln^2 \frac{Q^2}{\mu_r^2}}\nonumber \\
    &\times \int d^2k_\bot e^{ir_\bot \cdot k_\bot} \left\{ H_{\mathrm{Born}} f_1^\gamma(x, k_\bot^2) + H_{\mathrm{Born}}^{\cos(2\phi)} x h_1^{\perp \gamma}(x, k_\bot^2) \right\},
\end{align}
where
\begin{align}
     &H_{\mathrm{Born}} = 2\alpha_e^2 z^2 \left[ \frac{1 + (1-z)^2}{\left(P_\bot^2 + (1-z)Q^2\right)^2} - \frac{2Q^2 P_\bot^2 z^2 (1-z)}{\left(P_\bot^2 + (1-z)Q^2\right)^4} \right],
\\
&H_{\mathrm{Born}}^{\cos(2\phi)} = 2\alpha_e^2 z^2 \frac{-2Q^2 P_\bot^2 z^2 (1-z)}{\left(P_\bot^2 + (1-z)Q^2\right)^4} \cos{2\phi}.
\end{align}
\begin{figure}[htbp]
    \centering
    \begin{subfigure}{0.5\textwidth}
        \centering
        \includegraphics[width=\textwidth]{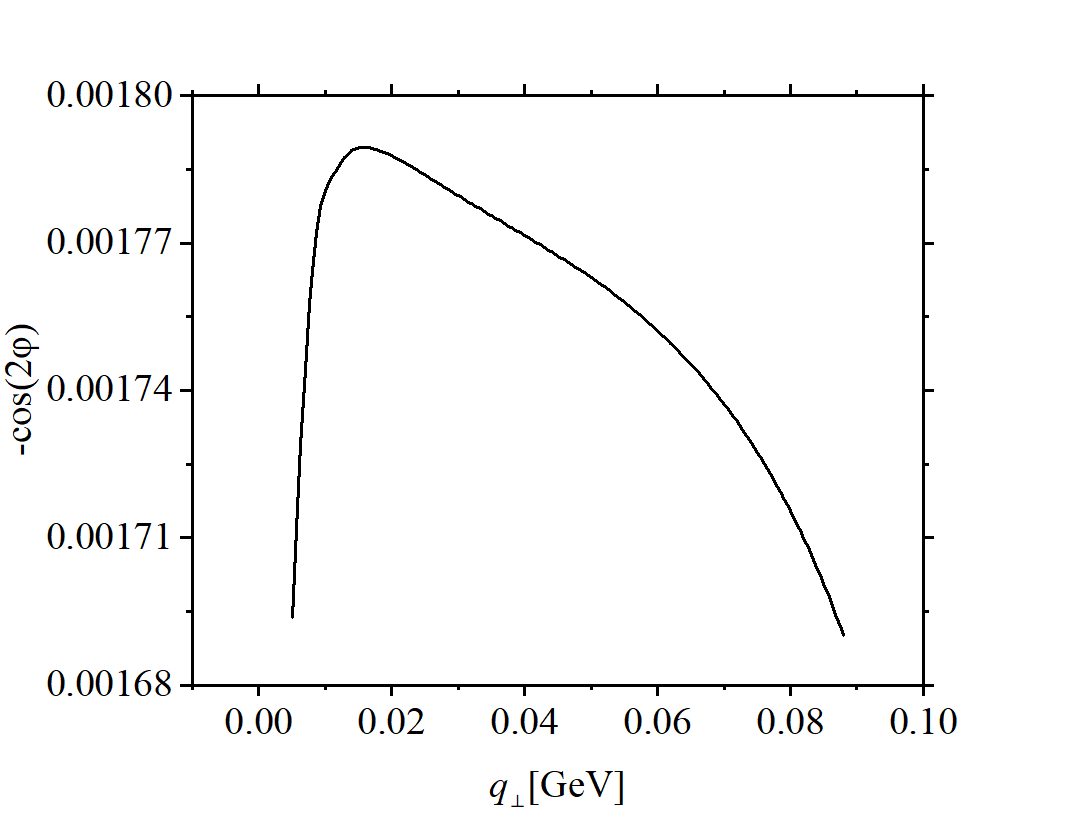} 
    \end{subfigure}\hfill
    \begin{subfigure}{0.5\textwidth}
        \centering
        \includegraphics[width=\textwidth]{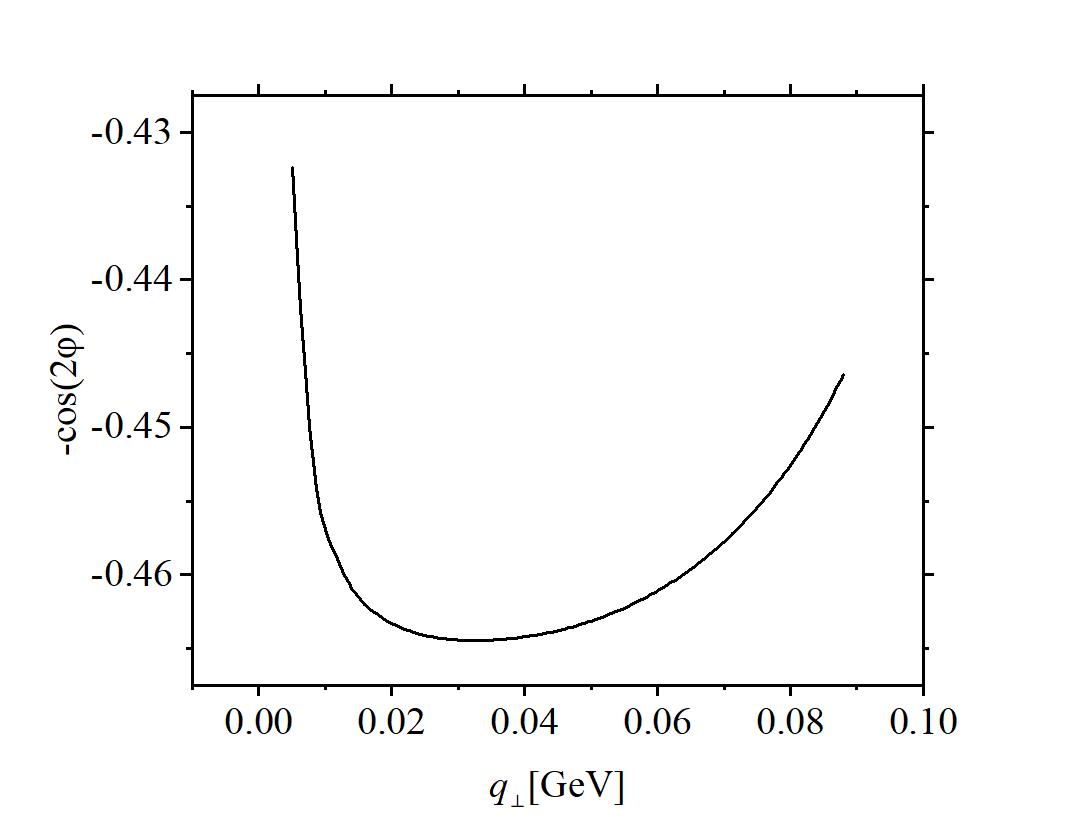} 
    \end{subfigure}
    \caption{The azimuthal asymmetry as the function of $q_{\perp}$ with (right panel) and without (left panel) taking into account the corrections of background field on coupling constant for a Pb target at EicC. $Q^2$ is fixed to be $Q^2=0.16\, \mathrm{GeV}^2$. The $P_{\perp}$ is integrated over the  region $[0.3 \,\mathrm{GeV}, 0.4\, \mathrm{GeV}]$. The emitted photon rapidity $y_\gamma$ is integrated over the region $[0.5,1]$.}
    \label{f.fig1}
\end{figure}
\begin{figure}[htbp]
    \centering
    \begin{subfigure}{0.5\textwidth}
        \centering
        \includegraphics[width=\textwidth]{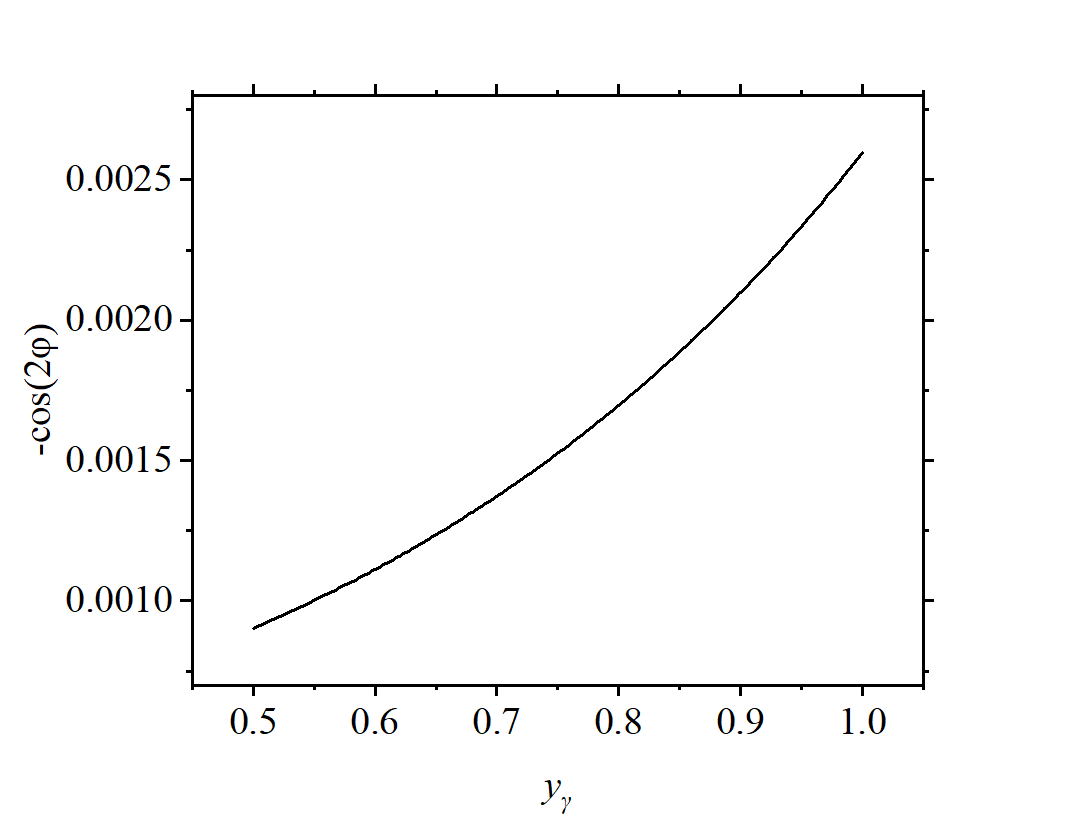} 
    \end{subfigure}\hfill
    \begin{subfigure}{0.5\textwidth}
        \centering
        \includegraphics[width=\textwidth]{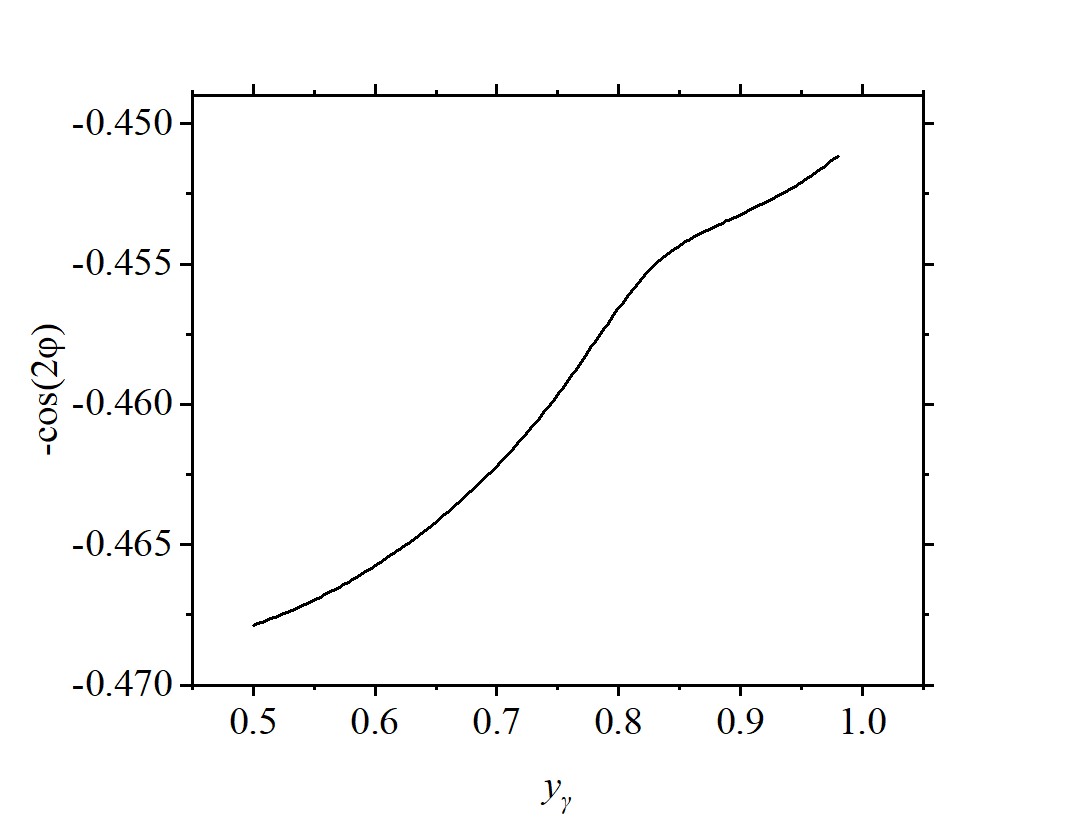} 
    \end{subfigure}
    \caption{The azimuthal asymmetry as the function of $q_{\perp}$ with (right panel) and without (left panel) taking into account the corrections of background field on coupling constant for a Pb target at EicC. $Q^2$ is fixed to be $Q^2=0.16 \mathrm{GeV}^2$. The $P_{\perp}$ is integrated over the  region $[0.3 \mathrm{GeV}, 0.4 \mathrm{GeV}]$, and the total transverse momentum $q_{\perp}$ is fixed to be 50 MeV.}
    \label{f.fig2}
\end{figure}

\begin{figure}[htbp]
    \centering
    \begin{subfigure}{0.5\textwidth}
        \centering
        \includegraphics[width=\textwidth]{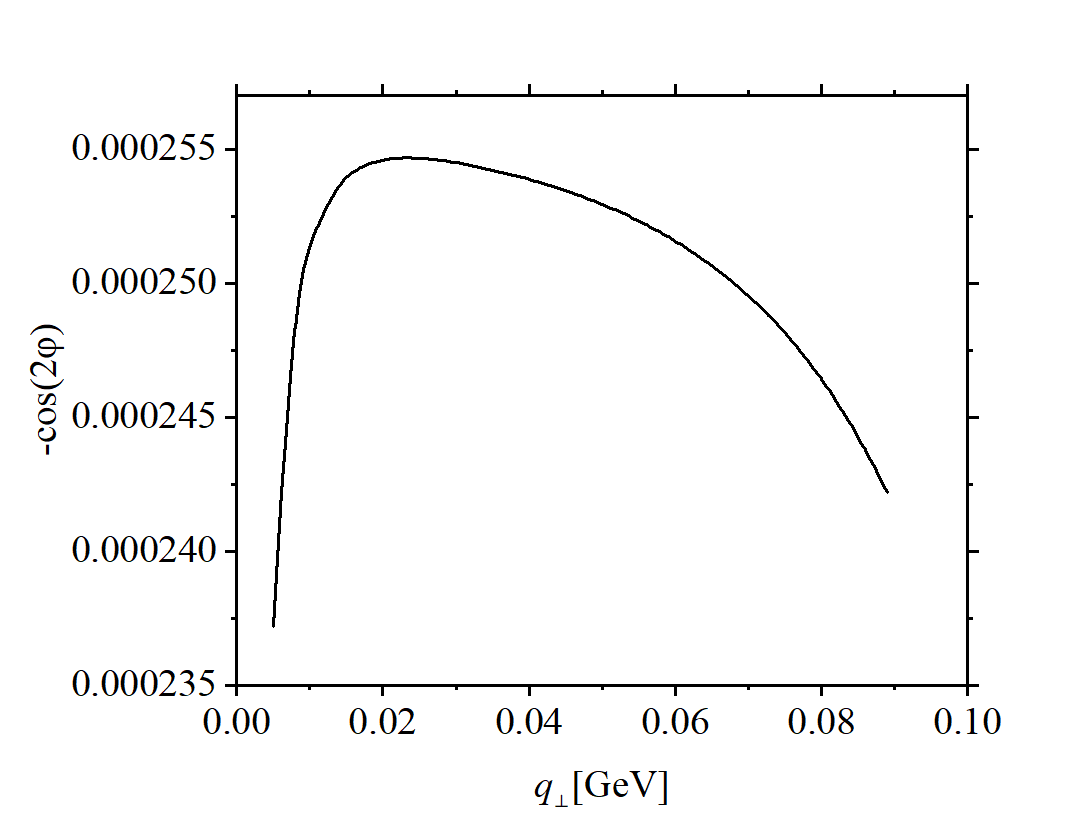} 
    \end{subfigure}\hfill
    \begin{subfigure}{0.5\textwidth}
        \centering
        \includegraphics[width=\textwidth]{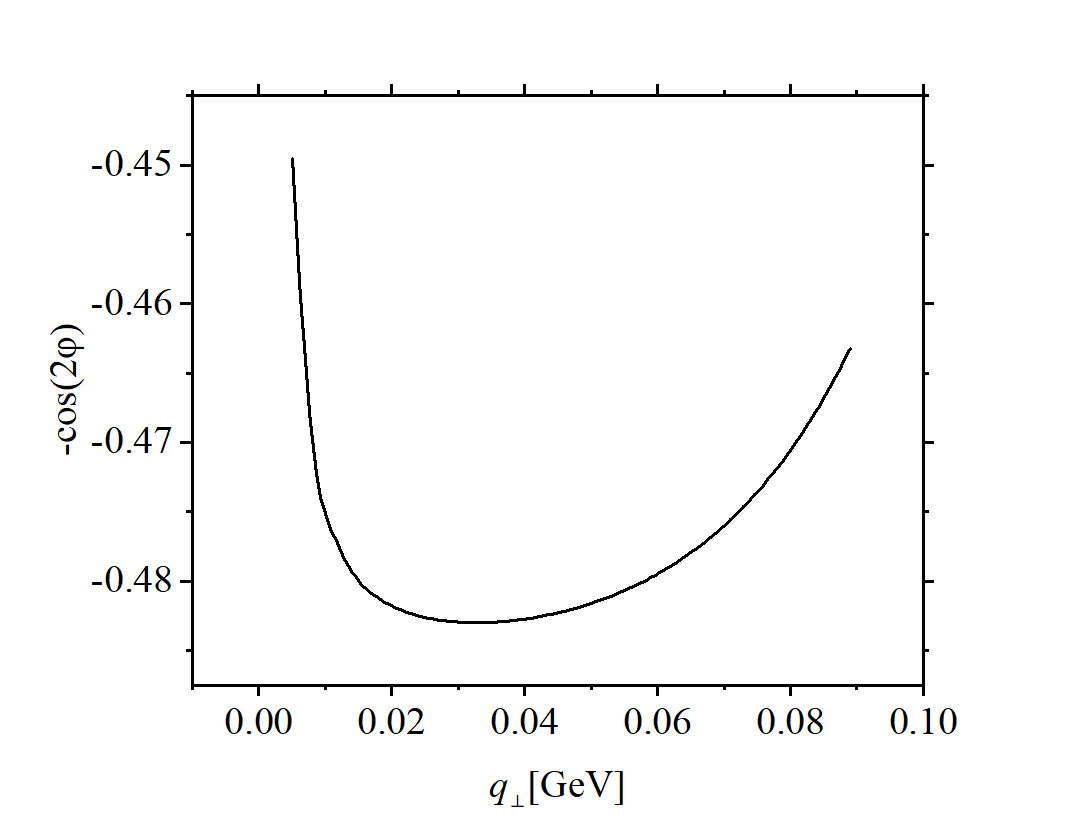} 
    \end{subfigure}
    \caption{The azimuthal asymmetry as the function of $q_{\perp}$ with (right panel) and without (left panel) taking into account the corrections of background field on coupling constant for a Pb target at EicC. $Q^2$ is fixed to be $Q^2=4\, \mathrm{GeV}^2$. The $P_{\perp}$ is integrated over the  region $[0.3 \,\mathrm{GeV}, 0.4 \,\mathrm{GeV}]$. The emitted photon rapidity $y_\gamma$ is integrated over the region $[0.5,1]$.}
    \label{f.fig3}
\end{figure}
\begin{figure}[htbp]
    \centering
    \begin{subfigure}{0.5\textwidth}
        \centering
        \includegraphics[width=\textwidth]{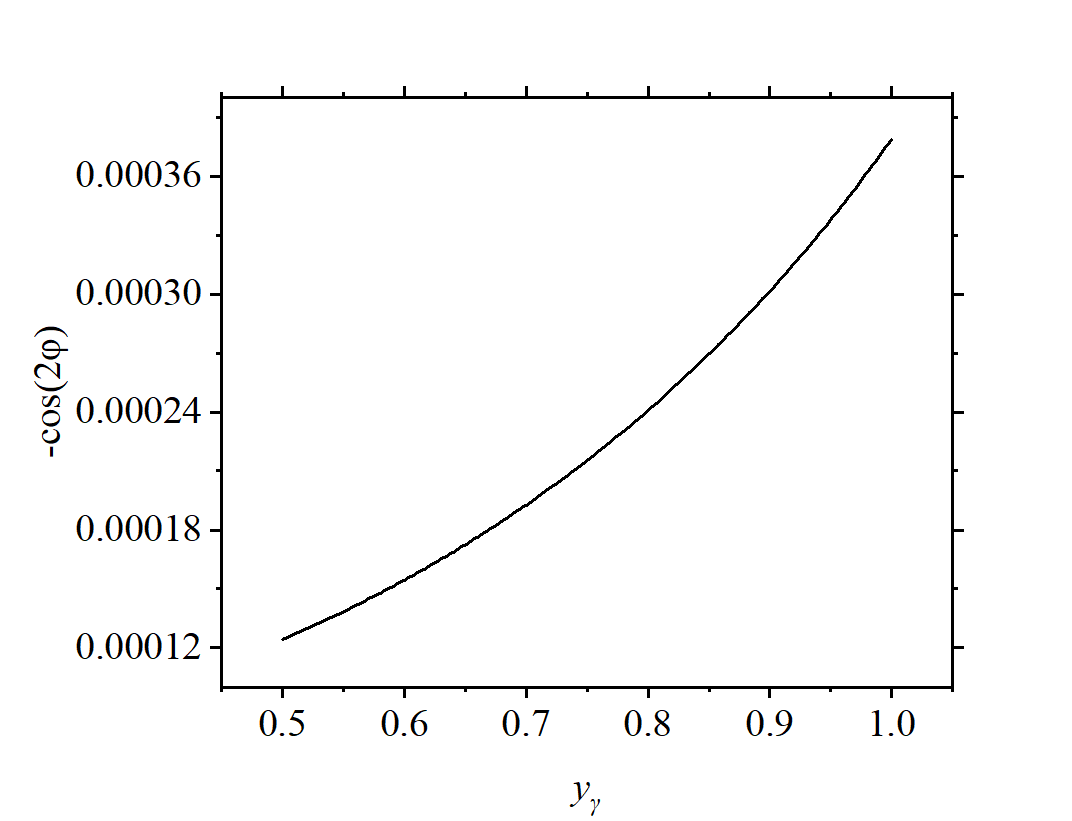} 
    \end{subfigure}\hfill
    \begin{subfigure}{0.5\textwidth}
        \centering
        \includegraphics[width=\textwidth]{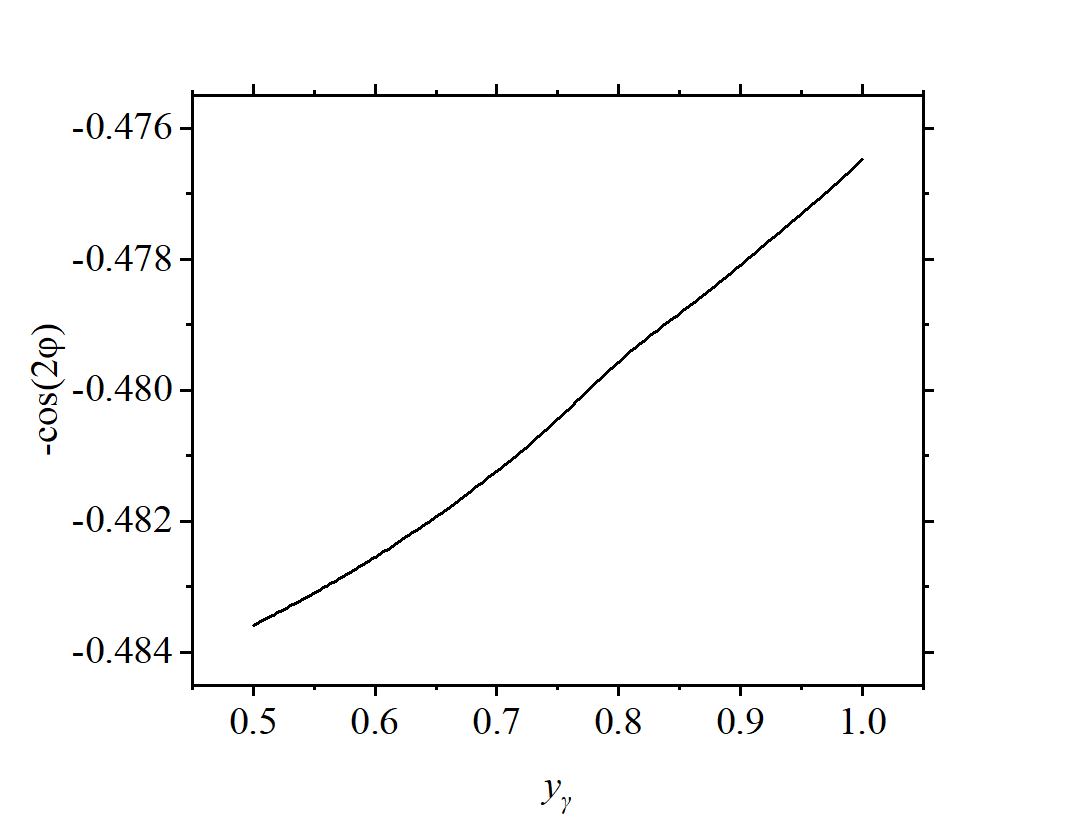} 
    \end{subfigure}
    \caption{The azimuthal asymmetry as the function of $q_{\perp}$ with (right panel) and without (left panel) taking into account the corrections of background field on coupling constant for a Pb target at EicC. $Q^2$ is fixed to be $Q^2=4 \,\mathrm{GeV}^2$. The $P_{\perp}$ is integrated over the region $[0.3\, \mathrm{GeV}, 0.4 \,\mathrm{GeV}]$, and the total transverse momentum $q_{\perp}$ is fixed to be 50 MeV.}
    \label{f.fig4}
\end{figure}
Now, we can numerically evaluate the magnitude of this effect at the EIC and EicC. We separately assess the dependence of the $\cos{2\phi}$ azimuthal asymmetry on the rapidity of the final-state photon and on the total transverse momentum at EIC and EicC energies.
For the EIC, the electron beam and heavy ion beam energies are 18 GeV and 100 GeV, respectively, while for the EicC, they are 3.5 GeV and 8 GeV, respectively. The average value of the azimuthal angle $\cos{2\phi}$ is given by
\begin{equation}
\left\langle\cos{\left(2\phi\right)}\right\rangle=\frac{\int\ \frac{d\sigma}{d\mathcal{P}.\mathcal{S}.}\cos{2\phi d\mathcal{P}}.\mathcal{S}.}{\int\ \frac{d\sigma}{d\mathcal{P}.\mathcal{S}.}d\mathcal{P}.\mathcal{S}.},
\end{equation}
where $d\mathcal{P}.\mathcal{S}$ is the phase space factor. 

\begin{figure}[htbp]
    \centering
    \begin{subfigure}{0.5\textwidth}
        \centering
        \includegraphics[width=\textwidth]{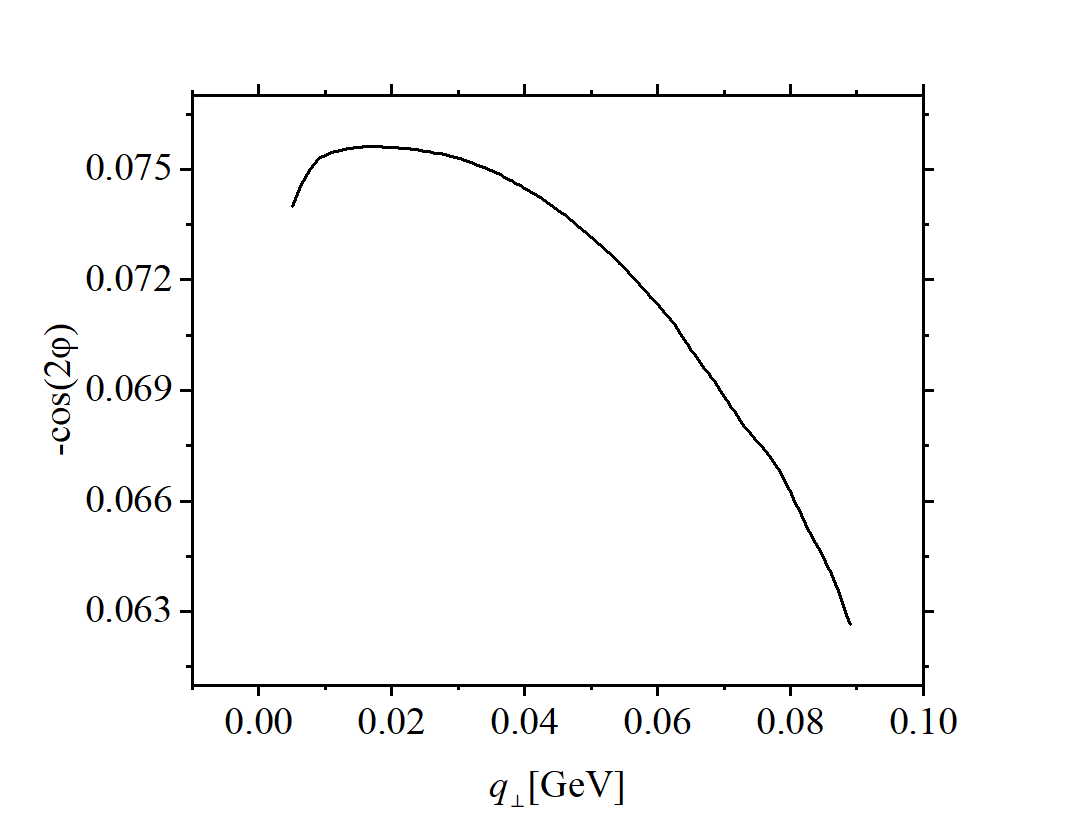} 
    \end{subfigure}\hfill
    \begin{subfigure}{0.5\textwidth}
        \centering
        \includegraphics[width=\textwidth]{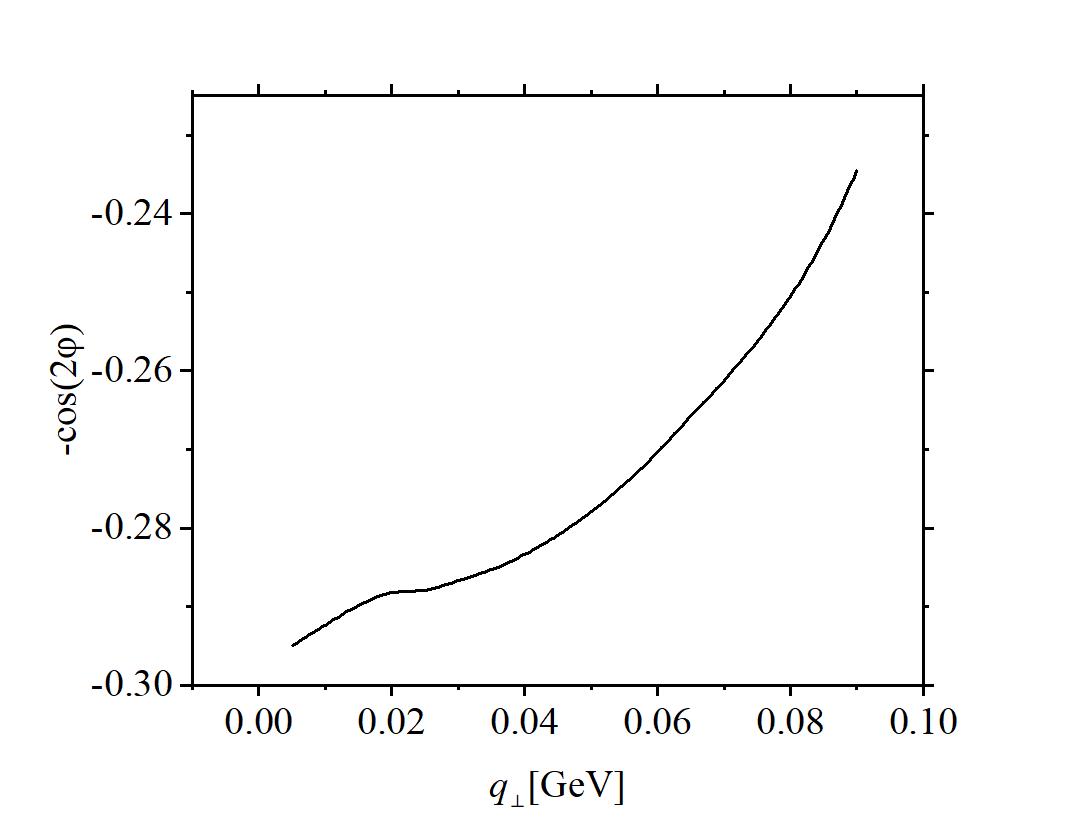} 
    \end{subfigure}
    \caption{The azimuthal asymmetry as the function of $q_{\perp}$ with (right panel) and without (left panel) taking into account the corrections of background field on coupling constant for a Pb target at EIC. $Q^2$ is fixed to be $Q^2=4\, \mathrm{GeV}^2$. The $P_{\perp}$ is integrated over the  region $[1.5\, \mathrm{GeV}, 2 \,\mathrm{GeV}]$. The emitted photon rapidity $y_\gamma$ is integrated over the region $[2,2.8]$.}
    \label{f.fig5}
\end{figure}

\begin{figure}[htbp]
    \centering
    \begin{subfigure}{0.5\textwidth}
        \centering
        \includegraphics[width=\textwidth]{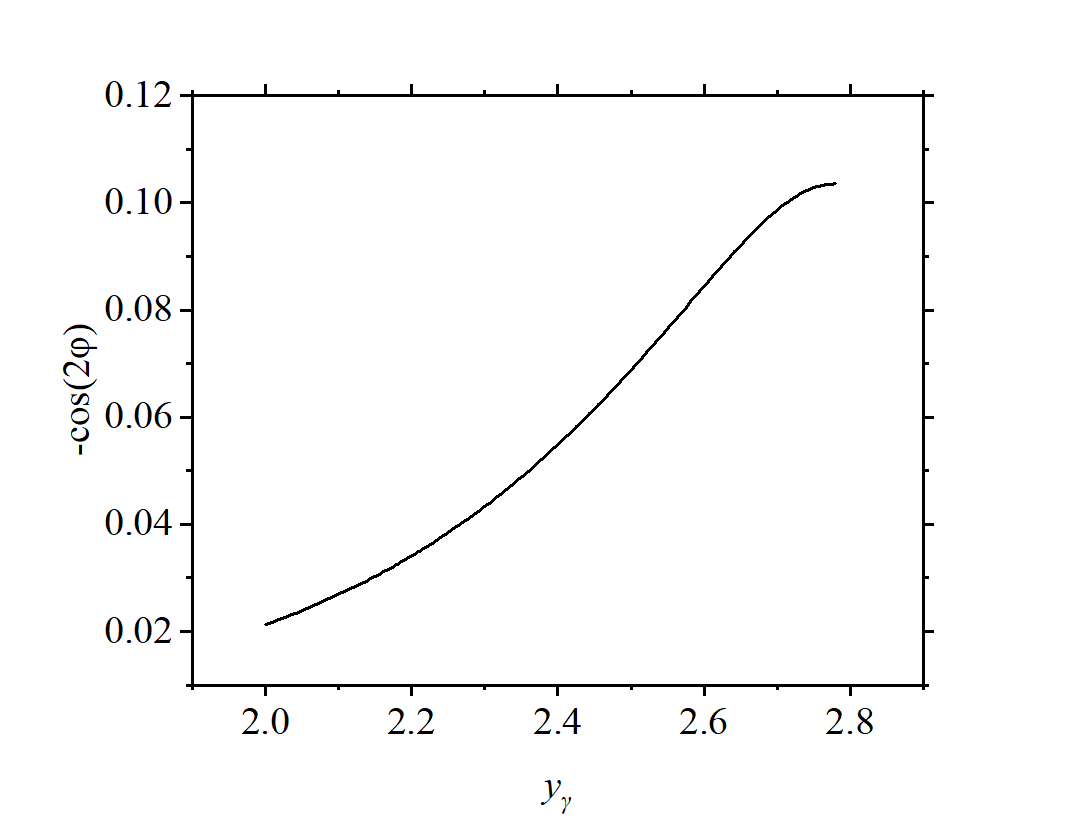} 
    \end{subfigure}\hfill
    \begin{subfigure}{0.5\textwidth}
        \centering
        \includegraphics[width=\textwidth]{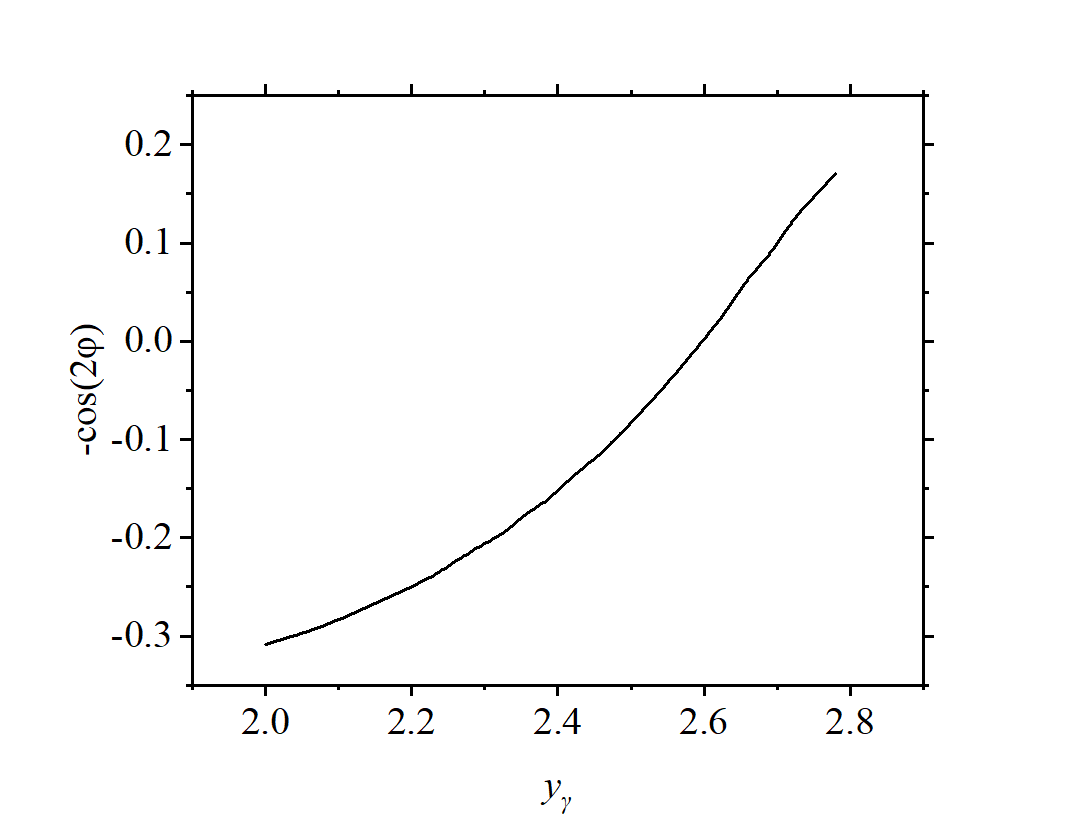} 
    \end{subfigure}
    \caption{The azimuthal asymmetry as the function of $q_{\perp}$ with (right panel) and without (left panel) taking into account the corrections of background field on coupling constant for a Pb target at EIC. $Q^2$ is fixed to be $Q^2=4\, \mathrm{GeV}^2$. The $P_{\perp}$ is integrated over the region $[1.5\, \mathrm{GeV}, 2 \,\mathrm{GeV}]$, and the total transverse momentum $q_{\perp}$ is fixed to be 50 MeV.}
    \label{f.fig6}
\end{figure}

In our numerical analysis, we examined three scenarios: first, $ Q^2 = 0.16 \, \text{GeV}^2 $ and $ Q^2 = 4 \, \text{GeV}^2 $ in EicC; second, $ Q^2 = 4 \, \text{GeV}^2 $ in EIC. In each scenario, we further explore the dependence of azimuthal asymmetry on the total transverse momentum $ q_\perp $ and the rapidity $ y $. Figs.~\ref{f.fig1},~\ref{f.fig3}, and~\ref{f.fig5} reveal significant differences in the dependence of $ q_\perp $ on azimuthal asymmetry before and after considering the correction. A notable observation is that, after the corrections, the asymmetry transitions to negative values and exhibits an increase in its absolute magnitude. Before corrections, the asymmetry initially increases with $ q_\perp $ up to a certain point before decreasing, while after corrections it follows a trend that consistently increases before eventually decreasing. 
Figs.~\ref{f.fig2},~\ref{f.fig4}, and~\ref{f.fig6} present the dependence of azimuthal asymmetry on rapidity $ y $, where substantial differences are observed before and after the corrections. The most significant change is once again the transition to negative asymmetry values, with the absolute magnitude becoming larger after corrections. In both cases, before and after corrections, the asymmetry increases monotonically with increasing $ y $. Furthermore, a comparison of Figs.~\ref{f.fig1} and~\ref{f.fig3}, as well as Figs.~\ref{f.fig2} and~\ref{f.fig4}, demonstrates that prior to the corrections, a smaller $ Q^2 $ value in EicC correlates with a larger asymmetry. However, after the corrections, the asymmetry no longer exhibits a significant dependence on $ Q^2 $. Finally, comparing Figs.~\ref{f.fig5} and~\ref{f.fig3}, along with Figs.~\ref{f.fig6} and~\ref{f.fig4}, indicates that within the relevant kinematic range of the EIC, the asymmetry is greater –both before and after corrections –under identical conditions $ Q^2 $.


\section{Summary}
\label{4}
In this paper, we have investigated the impact of polarized background fields on scattering processes, with a focus on a unique effect distinct from Coulomb corrections. While Coulomb corrections primarily account for the modification of the electron propagator by the background field, this study examines the influence of the background field on the photon propagator. By parameterizing the photon propagator within the polarized background field, we derive a corrected propagator, which can replace the standard propagator associated with final-state photon radiation.

To experimentally detect this effect, we propose measuring the azimuthal asymmetry of the final-state particles in the BH process at the EIC and EicC. Our numerical results demonstrate that this effect significantly influences the azimuthal asymmetry, making it favorable for experimental observation. We anticipate that this research will provide meaningful insights into understanding QCD processes in polarized gluon fields, particularly given the growing evidence from previous studies suggesting that gluon fields in hadrons exhibit polarization.

\section*{Acknowledgments}
AI tools were used to improve sentence fluency and make English more natural.

\bibliography{ref}

\end{document}